\documentclass[preprint]{chjaa}
\usepackage{graphicx,times}
\usepackage{indentfirst}
\usepackage{amsmath}
\usepackage{hyperref}
 \renewcommand{\vec}[1]{\boldsymbol{#1}}
 \renewcommand{\vec}[1]{\bm{#1}}

 \newcommand{\dif}{\mathrm{d}}
\begin{document}
\title{Microscopic magnetic dipole radiation in neutron stars}
%
\volnopage{Vol.0 (200x) No.0, 000--000}
%
\setcounter{page}{1}
\author{Hao Tong \inst{1} \mailto{} \and Qiu-he Peng \inst{1} \and Hua Bai\inst{1}}
\institute{Department of astronomy, Nanjing University, Nanjing, 210093, China \\%
           \email{htong\_2005@163.com}%
          }
\date{Received~~2001 month day; accepted~~2001~~month day}
\abstract{There is ${}^3P_2$ neutron superfluid region in NS
(neutron star) interior. For a rotating NS, the ${}^3P_2$ superfluid
region is like a system of rotating magnetic dipoles. It will give
out electromagnetic radiation, which may provides a new heating
mechanism of NSs. This heating mechanism plus some cooling agent may
give sound explanation to NS glitches.
}
\authorrunning{H. Tong \& Q. H. Peng \& H. Bai}
\titlerunning{MMDR in NS}
\maketitle %

\section{Introduction}

Since the discovery of NSs (neutron star) and glitches in late
1960s, it is generally believed that there is superfluidity in NS
interior (Ruderman 1976, Shapiro et al. 1983, Elgar{\o}y et al.
1996). Cooling mechanism associated with superfluidity was first
proposed by Flowers et al. (1976). Not until recently, the
importance of superfluidity is considered seriously in the "Minimal
model" (Gusakov et al. 2004, Page et al. 2004, Kaminker et al.
2006).

Then one may ask: since cooling agent associated with superfluidity
must be considered in the "Minimal model", what about the heating
mechanisms? Heating mechanism accompanied with superfluidity has
been discussed by Alpar et al. (1989), and was taken into NS cooling
model by Umeda et al (1994), Page et al. (2006), Tsuruta (2006).
Here we investigate another possibility: microscope magnetic dipole
radiation (MMDR) in NSs, possibly another heating mechanism
associated with superfluidity.

The idea is as follows: there is ${}^3P_2$ superfluid region in the
interior of a NS. We calculate its paramagnetic properties in the
presence of a background magnetic field. Since the neutron
superfluid is in a vortex state, the ${}^3P_2$ superfluid region is
like a system of rotating magnetic dipoles. This system will give
magnetic dipole radiation. The emitted photon can not penetrate the
NS matter, thus provides a heating mechanism of NSs. The origin of
this heating mechanism is microscopic magnetic dipole radiation, so
we call it MMDR heating of NSs.

This heating mechanism plus some cooling agent may give sound
explanation to NS glitches (Bai et al. 2006; Peng et al. 2006; Peng
2007).

Before going into detailed calculations of MMDR, we will first look
at the superfluidity in NSs.

\section{Superfluidity in NS}\label{superfluidity}

There are two relevant regimes of neutron superfluid inside the NS
interior: one is the isotropic ${}^1S_0$ neutron superfluid within a
density range $1\times 10^{10}< \rho\ (g\,cm^{-3})<1.6\times
10^{14}$. The critical temperature is $T_c({}^1S_0) \approx 1\times
10^{10}\,K$.

Another important regime is the anisotropic ${}^3P_2$ neutron
superfluid within a wide density range $1.3 \times 10^{14} < \rho\
(g\,cm^{-3}) < 7.2 \times 10^{14}$. The critical temperature is
\begin{equation}
T_c ({}^3P_2) = \Delta_{max}({}^3P_2)/2k \approx 2.78\times
10^{8}\,K.
\end{equation}
We note that the energy gap $\Delta({}^3P_2)$ is almost a constant
about the maximum with an error less than $3\%$ in a rather wide
density range $3.3\times 10^{14} < \rho\ (g\,cm^{-3}) < 5.2\times
10^{14}$ (see Fig.2 of Elgar{\o}y et al. 1996, but we neglect the
$F$ state of neutron Cooper pair here).

It is well known that a rotational superfluid must be in the
superfluid vortexes. In general the vortex filaments are arranged in
a symmetric lattice, they are parallel to the axis of rotation of NS
almost rigidly. The circulation of every vortex filament $\Gamma$ is
quantized
\begin{align}\label{Feynman}
\Gamma &=\oint \vec{v}\cdot \dif \vec{l} = n \Gamma_0,\\
\Gamma_0 &= \frac{2\pi \hbar}{2m_n}.
\end{align}
Here $n$ is a circulation quantum number of the vortex, $m_n$ is
neutron mass, $\hbar$ is the Planck's constant, $\Gamma_0$ is the
unit vortex quanta(Feynman 1955, Lifshitz et al. 1999).

It is supposed that the core of the superfluid vortex is a
cylindrical region of normal neutron fluid immersed in the
superfluid neutron sea. As usual, the radius of the core of vortex
$a_0$, is taken to be the coherent length of neutron superfluid
\begin{equation}
a_0 = \frac{E_F}{k_F\, \Delta} \approx (3\pi^2)^{1/3}
\frac{\hbar^2}{2\,m_n^{4/3}} \frac{\rho^{1/3}}{\Delta}.
\end{equation}
Here $E_F$ is the Fermi energy of the neutrons, $k_F$ is the
corresponding Fermi wave number, $\rho$ is the total density of the
neutron superfluid region(Ruderman 1976). Outside the core of the
vortex, neutrons are in a superfluid state.

The kinetic properties of superfluid vortexes follow directly from
the Feynman circulation theorem eq.\ref{Feynman}. The superfluid
neutrons revolve around the vortex line with a velocity
\begin{equation}
v_s(r) =\frac{n\,\hbar}{2m_n r},
\end{equation}
where $r$ is the distance from the field point to the axis of the
vortex filament. The distribution of the angular velocity of the
neutron superfluid revolving around the vortex filament is
\begin{equation}\label{omega}
\omega_s(r) = \frac{n\hbar}{2m_n r^2}.
\end{equation}
Therefore the revolution of the superfluid neutrons around the
vortex filament is placed in a differential state: the closer to the
center of the vortex, the faster the superfluid neutrons rotate.
Near $r\approx a_0$, the angular velocity reaches its maximum value
\begin{equation}
\omega_c = \frac{n\hbar}{2m_n a_0^2}.
\end{equation}
And inside the core of the vortex $r<a_0$, the normal neutron fluid
revolve rigidly at angular velocity of $\omega_c$.

According to Feynman (1955), the number of superfluid vortex
filament per unit area is $2\Omega/\bar{n}\Gamma_0$, where $\Omega$
is macroscopic angular velocity, $\bar{n}$ is mean circulation
quantum number. Then the average separation $b$ between vortex
filaments and the total number of the superfluid vortexes in the
interior of NS are respectively
\begin{align}
b &= \left ( \frac{\bar{n}\hbar}{2m_n \Omega} \right)^{1/2}\\
\label{vortex number} N_{vort} &= \left( \frac{R}{b} \right)^2 =
\frac{2m_n \Omega}{\bar{n}\hbar}\,R^2.
\end{align}
Here $R$ is the radius of the ${}^3P_2$ superfluid region.

In this paper, we only consider the thermodynamic condition. That is
$n=\bar{n}=1$. Although deviation from equilibrium have been
discussed by Yakovleve et al. (2001), Reisenegger et al.(1995,
2006).

For a ${}^3P_2$ neutron superfluid vortex in the interior of a NS,
$a_0 \sim 10^{-10}\,cm$ and $b\sim 10^{-3}\,cm$. Therefore the
vortex core is very tiny, the distribution of the vortex filament is
exceedingly sparse. The separation between vortex filaments reaches
a macroscopic scale, and superfluidity is just a macroscopic quantum
phenomenon.

The magnetic behavior of ${}^3P_2$ superfluid region will be
discussed in the next section.

\section{Induced paramagnetic moment of the ${}^3P_2$ neutron superfluid in the B-phase}

\subsection{Two phases of the ${}^3P_2$ neutron superfluid}

A ${}^3P_2$ neutron Cooper pair has spin angular momentum with a
spin quantum number, $s = 1$. The magnetic moment of the ${}^3P_2$
neutron Cooper pair is twice that of a neutron, $2\mu_n$ in
magnitude, where $\mu_n =  -0.966\times 10^{-23}\,erg\,gauss^{-1}$
is the anomalous neutron magnetic moment. And its projection on an
external magnetic field (z-direction) is $s_z \times 2\,\mu_n$, $s_z
=1,0,-1$. It is interesting to note that the behavior of the
${}^3P_2$ neutron superfluid is very similar to that of the liquid
${}^3He$ at very low temperature (Leggett 1975):
\begin{enumerate}
    \item The projection distribution for the magnetic moment of the ${}^3P_2$
neutron Cooper pairs in the absence of external magnetic field is
stochastic, or "Equal Spin Pair" (ESP) phase.  The ${}^3P_2$ neutron
superfluid is basically isotropic and with no significant magnetic
moment in the absence of  external magnetic field. We name it as the
A-phase of the ${}^3P_2$ neutron superfluid similar to the A-phase
of the liquid ${}^3He$ at very low temperature (Leggett 1975).
    \item However, the projection distribution for the magnetic moment of the
${}^3P_2$ neutron Cooper pairs in the presence of external magnetic
field is not stochastic. The number of the ${}^3P_2$ neutron Cooper
pair with paramagnetic moment is more than the ones with the
diamagnetic moment. Therefore, the ${}^3P_2$ neutron superfluid has
a net induced paramagnetic moment and its behavior is anisotropic in
the presence of external magnetic field. We name it as the B-phase
of the ${}^3P_2$ neutron superfluid similar to the B-phase of the
liquid ${}^3He$ at very low temperature (Leggett 1975).

\end{enumerate}

\subsection{Induced paramagnetic moment of the ${}^3P_2$ neutron superfluid in the B-phase}

We now consider the paramagnetism of ${}^3P_2$ neutron superfluid
regions. Following standard treatment of magnetism (Pathria 2003,
Feng et al 2005), the Hamiltonian of the system in the presence of
external field is
\begin{equation}
H = -2\vec{\mu}_n \cdot \vec{B} = -2\mu_{nz}B.
\end{equation}
Here $B$ is the external field (in the z direction), $2\vec{\mu}_n$
is the magnetic moment of the ${}^3P_2$ neutron Cooper pair,
$2\mu_{nz}$ is its projection on the z direction.

The ensemble average only gives an additional thermal factor
\begin{align}\label{thermal factor}
\langle 2\mu_n \rangle = 2\mu_n f(\mu_n B/k\,T)\\
f(\mu_n B/k\,T) = \frac{2 \sinh \beta 2 \mu_n B}{1 + 2\cosh \beta
2\mu_n B}.
\end{align}
In the limit of high temperature
\begin{equation}
f(\mu_n B/kT) = \frac{4}{3} \frac{\mu_n B}{kT}.
\end{equation}
The $\propto \frac{1}{T}$ behavior is just the Curie's law of
paramagnetism in terrestrial laboratory. The qualitative behavior is
: as a NS cools down, its internal temperature $T$ decrease, while
the thermal factor increase. The ${}^3P_2$ neutron Cooper tends to
align in the same direction. This is the mathematica formalism of
the B-phase of the ${}^3P_2$ neutron superfluid.

As mentioned by Lifshitz et al. (1999), there is a finite
probability for two neutrons to combine into a Cooper pair. Since
only particles in the vicinity of the Fermi surface contributes,
only a finite fraction $q$ of the Fermi sphere are in the condensate
state
\begin{equation}
\begin{split}
q &= \frac{4\pi p_F^2 \Delta k}{\frac{4\pi}{3} p_F^3}\\
  &=3\sqrt{\frac{\Delta}{E_F}}\\
  &\sim 0.087.
\end{split}
\end{equation}
We have used the relation: $\Delta k = \sqrt{2m_n \Delta}, p_F =
\sqrt{2m_n E_F}$. Here $\Delta \sim 0.05\,MeV$ is the energy gap of
the corresponding ${}^3P_2$ superfluid region, $E_F \sim 60\,MeV$ is
the Fermi energy of the neutron system, $p_F$ is the corresponding
Fermi momentum, $\Delta k$ is the thickness of the shell which will
combine to Cooper pairs.

In conclusion, for a specific volume $\Delta V$ of ${}^3P_2$ neutron
superfluid region, only a small fraction of the Fermi sphere can
combine into ${}^3P_2$ Cooper pairs. Ensemble average gives another
thermal factor. So the net magnetic moment of the specific volume is
\begin{equation}
M_{\Delta V} = \Delta V \rho N_A \mu_n \, q \, f(\mu_n B/kT).
\end{equation}
Here $N_A$ is Avogadro's constant. The volume $\Delta V$ is
specially chosen so that it is large enough to contain an numerous
number of Cooper pairs, but small enough compared with macroscopic
scale. It is a mesoscopic volume. This is required by the coherence
calculation of the microscopic magnetic dipole radiation (MMDR)
below.

For a rotating NS, the ${}^3P_2$ superfluid region is in vortex
state. It have paramagnetic moment in the presence of external
field, on the mesoscopic scale. So the ${}^3P_2$ neutron superfluid
vortex is like a system of rotating magnetic dipoles, which will
give magnetic dipole radiation. This radiation can not penetrate the
NS matter, thus provides a heating mechanism of NS associate with
superfluidity.

The detailed calculation of MMDR is the scope of the next section.

\section{Microscopic magnetic dipole radiation heating}

\subsection{A working assumption}

In a superpluid vortex, each superfluid neutron revolve round the
axes of the vortex with angular velocity $\omega(r)$
(eq.\ref{omega}). It is well known that a rotating magnetic dipole
will give out magnetic dipole radiation (MDR). Therefore, the
${}^3P_2$ superfluid neutron will give electromagnetic radiation.
The frequency of emitted photons is equal to the rotational
frequency of the neutron.

We will use a phenomenological method to explore the radiation
problem in this paper (see, e.g. Feynman 1955; Androdikashvili et
al. 1966). The process is as follows: the rotational velocities of
superfluid neutron will decrease during the emission of the magnetic
dipole radiation due to the dissipation of its energy. The
superfluid neutrons, hence, will drift out according to
eq.\ref{omega}, and then the transverse pressure exerting on the
normal neutrons in the vortex cores will decrease. The normal
neutrons will move out to $r > a_0$ and become superfluid ones. At
the end of the vortexes, other normal neutrons located in the normal
neutron layer in the interior of neutron stars will flow into vortex
cores along the axes. At the same time the densities of superfluid
neutrons at the boundaries of vortex lattices will increase driving
a flow into the normal neutron layer along the direction
perpendicular to the axes. A "local circulation" will be formed in
the superfluid vortex region in this way. This process is very
similar to the  Ekman pumping (Greenspan 1968, Anderson et al.
1978). The normal neutrons in the crust of the neutron star will
drift inward during the process of these local circulations, and the
crust will slowly shrink a little. We may suppose that the energy of
the magnetic dipole radiation is really transformed by the released
gravitational energy of the inward moved crust through the local
circulation.

\subsection{Microscopic magnetic dipole radiation heating}

First, we calculate the magnetic dipole radiation by one single
vortex. The power radiated by one neutron is (Huang et al. 1982)
\begin{align}
W(n) &= \frac{2\omega^4}{3c^3} \sum_f |\langle f| \hat{M}_z |i
\rangle|^2\\ \nonumber
     &= \frac{2\omega^4}{3c^3} \sum_f \langle i| \hat{M}_z^{\dag} |f
\rangle \langle f | \hat{M}_z | i \rangle \\
     & =\frac{2\omega^4}{3c^3} \langle i | \hat{M}_z^{\dag} \hat{M}_z
| i \rangle .
\end{align}
where $c$ is the speed of light, $\omega$ is the angular velocity of
${}^3P_2$ superfluid neutron, $|f\rangle$ is the final state,
$|i\rangle$ is the initial state, and $\hat{M}_z$ is the operator of
magnetic moment. If we consider a coherent small volume $\Delta V$,
and take
\begin{equation}\label{magnetic moment}
\langle i |\hat{M}_z^{\dag} \hat{M}_z |i \rangle = \langle i |
\hat{M}_z^2 |i \rangle \approx M_{\Delta V}^2 \sin^2 \theta.
\end{equation}
Here $M_{\Delta V}$ is the corresponding paramagnetic moment in a
specific volume of $\Delta V$, $\theta $ is the angle between the
background field $\vec{B}$ and the rotational axis $\vec{\Omega}$.
Thus the required magnetic dipole radiation rate is
\begin{equation}\label{power }
W_{\Delta V} = \frac{2\omega^4}{3c^3} |M_{\Delta V}|^2 \, \sin^2
\theta,
\end{equation}
The formula is similar to the classical case (Shapiro et al. 1983),
because we have made approximations in obtaining eq.\ref{magnetic
moment}.

In calculation of the radiation power, the coherent effect must be
taken into consideration (Peng et al. 1982). This is like the
coherent effect in electromagnetism. We first calculate the
contribution of a specific volume $\Delta V$, then make a summation
of the all the specific volumes. This is done for one superfluid
vortex. The total power of MMDR is simply the sum of all single
vortex contributions. See the appendix for details.

The total power of microscopic magnetic dipole radiation (MMDR) is
\begin{equation}
W_{tot} = N_{vort}\, \eta\, A \frac12 \log \frac{b}{a_0}.
\end{equation}
Here $N_{vort}$ is total number of superfluid vortexes
eq.\ref{vortex number}, $\eta$ is the efficiency of coherence.
Radiation of one single vortex is $A \frac12 \log \frac{b}{a_0}$,
where the contribution factor $A$ is
\begin{equation}
A = \frac{8 \pi^4}{3c^2} \sin^2 \theta |\rho N_A \mu_n q f|^2 R
\omega_c^3 a_0^4.
\end{equation}
The factor $A$ is proportional to $R$, the vortex number $N_{vort}$
is proportional to $R^2$, as can be seen in eq.\ref{vortex number}.
So the total power $W_{tot}$ is proportional to $R^3$, that is
proportional to the volume of the ${}^3P_2$ neutron superfluid
region. This is what it should be.

The essential and discussion of MMDR heating is presented in the
next two sections.

\section{MMDR vs. other heating mechanisms}

Heating mechanisms in NSs can be classified into different
categories according to their energy input. There are several kinds
of energy input: magnetic, rotational, chemical, and confinement
energy etc. The corresponding heating mechanisms are respectively:
\begin{enumerate}
    \item Ohm heating (Page et al. 2006 and reference therein);

    \item Vortex creep heating (Alpar et al. 1989, Umeda et al. 1994);

    \item Retochemical heating (Reisenegger 1995, Reisenegger et al.
    2006).

    \item Deconfinement heating (Yuan et al. 1999, Kang et al. 2007);

\end{enumerate}

MMDR heating has two distinguished points compared with the heating
mechanisms stated above:
\begin{enumerate}
    \item It is a heating mechanism associated with superfluidity,
    thus no superfluid supression.
    \item Its energy input is gravitational energy as stated in our
    working assumption.
\end{enumerate}

\section{Conclusions}

Here we have presented another possible heating mechanism of NSs,
associated with superfluidity. It can be compared with other heating
mechanisms and  cooling agents of NSs (Gusakov et al. 2004, Page et
al. 2006).
\begin{figure}[!htbp]
\centering
  \includegraphics[width=\textwidth]{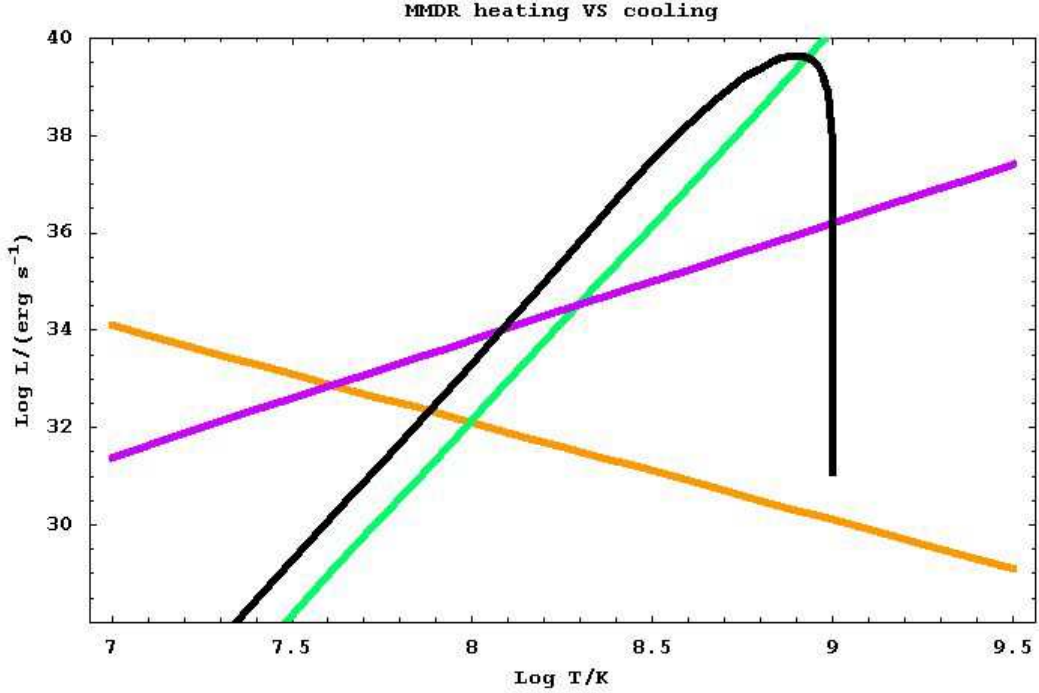}\\
  \caption{MMDR Heating VS Cooling. The brown one represents MMDR heating.
  It increase with decreasing temperature, as a result of increasing thermal factor.
  The red, green, and black one corresponds to photon cooling,
  MUrca (Modified Urca) process, and PBF (Pair Breaking and Formation) process
  respectively. (Adapted from Gusakov et al. (2004).)}
  \label{MMDR}
\end{figure}

As shown in Fig. \ref{MMDR}, MMDR heating may be dominate only in
the photon cooling stage. So it will not affect the cooling scenario
of young and mid age NSs. But for old NSs, e.g. PSR 1055-52, it may
serve as a moderate heating (Tsurata 2006, Page 2006).

Using toy model given by Yakovlev et al. (2003), we can make
illustrative calculation of NS cooling, including the MMDR heating.
\begin{figure}[!htbp]
\centering
  \includegraphics[width=\textwidth]{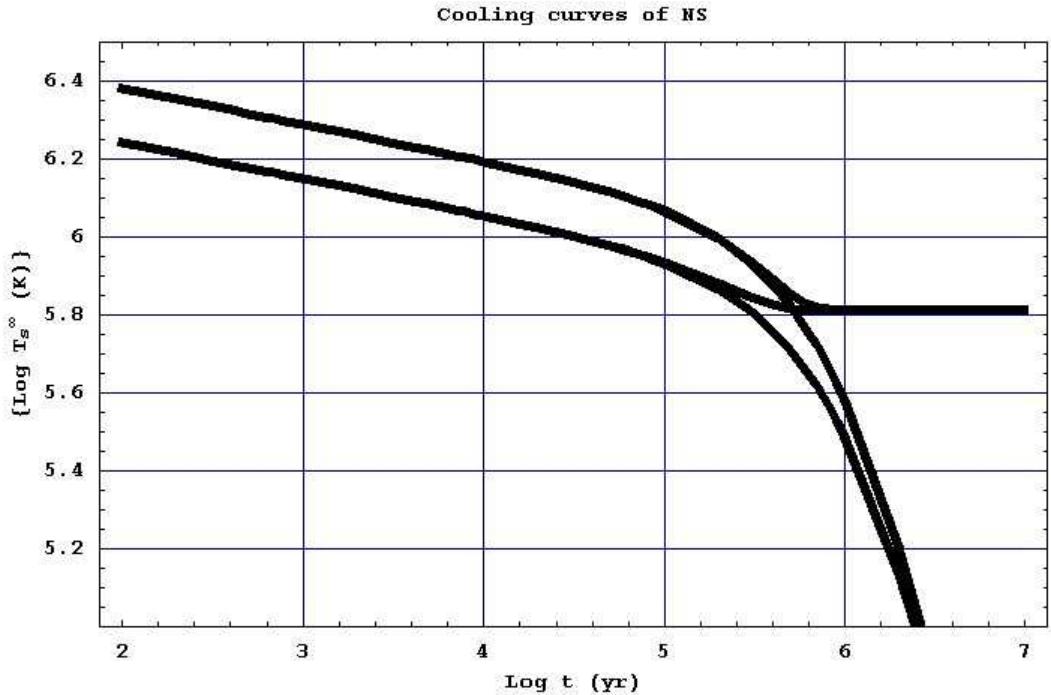}\\
  \caption{Cooling Curves Including MMDR Heating.
  The lower and upper curve correspond to MUrca and bremsstrahlung process
  dominated case respectively.
  Whereas the plateau at latter stage is due to MMDR heating.
  The physical input are the same as Yakovlev et al. (2003) except that
  we include the MMDR heating.
  Remind: these cooling curves are for illustrative use only.}
  \label{CoolingCurves}
\end{figure}
Fig. \ref{CoolingCurves} shows that, there is a plateau in the late
stage of NS cooling curves. This plateau can be compared with that
of Kang's (2007).

This MMDR heating must cease after some time. In our case, the cease
of MMDR heating has several possibilities.
\begin{enumerate}
    \item As shown by Huang et al. (1982), the energy input of MMDR
    heating is through Ekman pumping. When this agent is out of
    work, MMDR heating have to make a pause.
    \item When there is a phase transition in the core, e.g.
    deconfinement of hardrons, all the hadron processes disappear including MMDR
    heating.
    \item Due complexities of superfluidity gap (Elagr{\o}y et al. 1996), the
    ${}^3P_2$ superfluidity region becomes slimer when the core becomes more
    compact. The MMDR heating contribution is ignorable in this case.
\end{enumerate}

We presented another possible NS heating mechanism associated with
superfluidity. The exact effect of MMDR heating needs accurate
calculation of the cooling curves. This heating mechanism plus some
cooling agent may give sound explanation to NS glitches. The
detailed investigation is the scope of another work (Peng 2007).

\section*{acknowledgements}

This research is supported by Chinese National Science Foundation
No.10573011, No.10273006, and the Doctoral Program Foundation of
State Education Commission of China.

\appendix

\section{Detailed calculation of MMDR power}

We will follow the routine of Peng et al. (1982). At radius $r$ from
the center of ${}^3P_2$ neutron superfluid vortex, the neutron
rotational frequency is
\begin{equation}
\omega_s(r) = \frac{\hbar}{2m_n r^2}.
\end{equation}
As stated in the main text, the frequency of radiated photons is
equal to the neutron rotational frequency  $\omega_s(r)$. The wave
length of the radiated photon is
\begin{equation}\label{lambda}
\lambda_s(r) = \frac{2\pi c}{\omega_s(r)} = \frac{4\pi c m_n
r^2}{\hbar} \quad (r>a_0).
\end{equation}
The length of one single superfluid vortex is $\bar{H} =
\frac{\pi}{2} R$, where $R$ is the radius of the superfluid region.
One superfluid vortex can be separated into several segments, we
need only consider the coherent effect inside each segment. If we
assume the a length scale $\lambda_s(r)$, then a cylindrical shell
between the radius $r$ and $r+\dif r$ can be cut into
$\bar{H}/\lambda_s(r)$ segments. Here the specific volume $\Delta V$
is
\begin{equation}
\Delta V = 2\pi r \dif r \lambda_s(r).
\end{equation}
Paramagnetic moment in the specific volume is
\begin{equation}
M_{\Delta V} = \Delta V \rho N_A \mu_n q f\left( \frac{\mu_n B}{kT}
\right).
\end{equation}
Using eq.\ref{power }, differential power of one single superfluid
vortex is
\begin{align}
\dif W^{(1)} &= \frac{2\omega^4}{3c^3} \frac{\bar{H}}{\lambda_s(r)}
           |M_{\Delta V}|^2 \sin^2 \theta \\
        &= \frac{2\omega^4}{3c^3} \frac{\bar{H}}{\lambda_s(r)}
        \sin^2 \theta |\rho N_A \mu_n q f|^2 |2\pi r \dif r
        \lambda_s(r)|^2\\
        &= \frac{8\pi^2 \omega^4}{3c^3} \bar{H} \lambda_s(r) \sin^2\theta |\rho N_A \mu_n q
        f|^2 r^2 \dif r^2 \\
        &= \frac{8\pi^4}{3c^2} \sin^2\theta |\rho N_A \mu_n q f|^2 R
        \omega^3 r^2 \dif r^2.
\end{align}
In obtaining the final expression, we have used the definition of
$\lambda_s(r)$ and $\bar{H}$. Using non-dimensional quantities
\begin{equation}
\begin{split}
\omega_s(r) &=\omega_c \omega^{\prime}(r)\\
r &= a_0 r^{\prime}\\
\omega^{\prime} &= \frac{1}{r^{\prime^2}},
\end{split}
\end{equation}
the differential power is
\begin{align}
\dif W^{(1)} &=\frac{8\pi^4}{3c^2} \sin^2\theta |\rho N_A \mu_n q
f|^2 R \omega_c^3 a_0^4 \omega^{\prime 3} r^{\prime 2} \dif
r^{\prime 2}\\
             &= A \omega^{\prime 3} r^{\prime 2} \dif
r^{\prime 2}\\
             &= A \dif I
\end{align}
When integrating $\dif I$, a $\delta$-function is included
automatically, $\delta(\omega^{\prime}-1/r^{\prime 2})$. Performing
the integration from $a_0$ to $b$ gives $\frac12 \log
\frac{b}{a_0}$. The contribution factor $A$ is
\begin{equation}
A = \frac{8\pi^4}{3c^2} \sin^2\theta |\rho N_A \mu_n q f|^2 R
\omega_c^3 a_0^4.
\end{equation}
In the end, the microscopic magnetic dipole radiation power of one
superfluid vortex is
\begin{equation}
W^{(1)} = A \frac12 \log \frac{b}{a_0}.
\end{equation}

In the above calculation, we assumed a length scale $\lambda_s(r)$.
More generally, if the length scale is taken to be $\eta
\lambda_s(r)$, where $\eta$ is an efficiency factor, the MMDR power
of one superfluid vortex is
\begin{equation}
W^{(1)} = \eta A \frac12 \log \frac{b}{a_0}.
\end{equation}
Discussions in Sect.\ref{superfluidity} provides the total number of
superfluid vortexes
\begin{equation}
N_{vort} = \left( \frac{R}{b} \right)^2 = \frac{2m_n
\Omega}{\hbar}\,R^2.
\end{equation}
In conclusion, the total power of MMDR in the ${}^3P_2$ neutron
superfluid region in the interior of NS is
\begin{equation}
W_{tot} = N_{vort} \eta A \frac12 \log \frac{b}{a_0}.
\end{equation}
This is what we use in the main text.

\label{lastpage}
\end{document}